\documentclass[11pt,a4paper]{article}
\usepackage[T1]{fontenc}
\usepackage{lmodern}
\usepackage[margin=25mm]{geometry}
\usepackage{amsmath,amssymb,graphicx,booktabs}
\usepackage{microtype}
\usepackage[numbers,sort&compress]{natbib}
\usepackage{xurl}
\usepackage[hidelinks]{hyperref}
\hypersetup{pdftitle={Robustness and Performance of Linear Solvers for Cardiovascular Fluid-Structure Interaction Simulations},pdfauthor={Zheyi Chu and Chi Zhu}}

\title{Robustness and Performance of Linear Solvers for Cardiovascular Fluid-Structure Interaction Simulations}
\author{Zheyi Chu and Chi Zhu\thanks{Corresponding author.
  \href{mailto:chi.zhu@pku.edu.cn}{\nolinkurl{chi.zhu@pku.edu.cn}}.
  ORCID \url{https://orcid.org/0000-0002-1099-8893}.}\\[4pt]
  \small School of Mechanics and Engineering Science, Peking University\\
  \small Beijing 100871, China}
\date{}

\begin{document}
\maketitle

\begin{abstract}
The computational cost of monolithic cardiovascular fluid-structure interaction (FSI) simulations is dominated by the solution of large, tightly coupled linear systems. This study systematically investigates the robustness and efficiency of Krylov solvers with algebraic and physics-based preconditioners in an open-source monolithic arbitrary Lagrangian-Eulerian framework, with particular emphasis on how structural formulation alters the block structure and solver behavior. The numerical experiments cover a cylindrical pipe and three patient-specific models using both velocity- and displacement-based structural formulations. Results show that solver performance depends strongly on both problem configuration and structural formulation, with the displacement-based formulation substantially reducing the robustness of several preconditioners. Block-level spectral analysis links this behavior to severe ill-conditioning of the isolated pressure block. By retaining the velocity-pressure coupling and improving the conditioning of the reduced pressure problem, Schur-complement preconditioning provides the most consistent convergence across the investigated cases. A non-nested lower-triangular factorization with BoomerAMG approximations of both block inverse actions provides the best balance between robustness and computational cost among the configurations tested. These findings provide a practical basis for preconditioner design and demonstrate the importance of accounting for formulation-induced block structure in cardiovascular FSI.
\end{abstract}


\section{Introduction}
\label{sec:intro}

Cardiovascular fluid-structure interaction (FSI) models resolve the interaction between blood flow and deformable vessel walls and are increasingly used to investigate patient-specific hemodynamics \citep{torii2006fluid,bazilevs2010fully}. Monolithic arbitrary Lagrangian-Eulerian (ALE) formulations are attractive for these problems because they enforce fluid-solid coupling within a single nonlinear system and avoid the stability restrictions associated with loosely coupled partitioned schemes \citep{causin2005}. This tight coupling, however, comes at substantial computational cost. After discretization and linearization, each Newton iteration requires the solution of a large, sparse, nonsymmetric, and indefinite linear system.

The difficulty of these linear systems is not governed by problem size alone. Fluid incompressibility introduces strong velocity-pressure coupling, while wall deformability increases the matrix bandwidth and weakens diagonal dominance \citep{Jongming1}. Their algebraic properties also depend on vascular geometry, flow regime, physiological boundary conditions, time-step size, structural constitutive response, and the formulation used for the vessel wall. Consequently, solver performance can vary substantially across cardiovascular FSI problems, and a configuration that is efficient and robust for one case may perform poorly for another. Robust solver design therefore requires understanding both how solver performance transfers across representative FSI problems and how modeling choices modify the resulting linear systems.

Of particular interest in this study is the structural formulation. Solid mechanics may be written either with displacement as the primary structural variable or in a first-order velocity-pressure form \citep{scovazzi2016simple}. Although both formulations describe the same structural dynamics at the continuous level, their discretized and linearized systems are not algebraically identical. Structural tangent contributions, stabilization terms, pressure variables, and interface couplings enter the global system differently. These differences can modify the conditioning and coupling of the individual blocks and, consequently, the effectiveness of linear solver configurations.

These linear systems are usually solved with Krylov subspace methods such as GMRES and BiCGStab \citep{saad1986gmres,van1992bicgs}, whose convergence depends strongly on the quality of the preconditioner \citep{benzi2002preconditioning,Benzi_Golub_Liesen_2005}. General algebraic preconditioners act directly on the assembled matrix. Incomplete LU factorizations, often combined with additive Schwarz domain decomposition for parallel execution, provide local approximations whose effectiveness depends on fill level, overlap, ordering, and subdomain structure \citep{meijerink1977ilu,saad1994ilut,cai1999restrictedASM}. Algebraic multigrid (AMG) methods seek scalability through coarse-space hierarchies and are highly effective for many elliptic operators. Their performance on monolithic FSI systems, however, depends on how well the multigrid components represent the distinct but strongly coupled fluid and structural operators \citep{vanvek1996algebraic}.

Physics-based preconditioners instead exploit the block structure induced by the governing equations. For the present monolithic FSI formulation, the linearized system is organized as
\begin{equation}
\label{eq:block-mat}
A=
\begin{bmatrix}
K & G \\
D & L
\end{bmatrix},
\end{equation}
where $K$ contains the momentum contributions from both the fluid and solid domains, $G$ and $D$ describe the velocity-pressure coupling, and $L$ contains pressure-related contributions, including those introduced by stabilization \citep{esmaily2013new}. The numerical properties and composition of these blocks depend on the structural formulation. Simple block splittings (additive or multiplicative) retain limited coupling information, whereas Schur-complement preconditioning incorporates the reduced pressure operator $S=L-DK^{-1}G$ and therefore more directly represents the coupled velocity-pressure response \citep{benzi2002preconditioning}.

Schur-complement preconditioning is well established for incompressible flow. Pressure convection-diffusion and least-squares commutator methods provide scalable approximations of the reduced pressure operator \citep{PCD,LSC}, although their effectiveness can depend on the discretization and stabilization \citep{LIU-nest}. Related approaches include the bi-partitioned iterative method, which constructs coupled velocity and pressure Krylov spaces \citep{BIPN-2015}, and nested block preconditioners, which approximate the block inverse through inner Krylov-AMG solves within an outer flexible GMRES iteration \citep{LIU-nest,saad1993flexible}. Although these preconditioners can be applied directly to monolithic FSI systems, their favorable performance for incompressible flow may not carry over because the inclusion of structural dynamics modifies the coupled block operators. Moreover, the nature and extent of these modifications depend on the structural formulation. A substantial literature has also addressed block preconditioning for monolithic FSI systems \citep{heil2004efficient,gee2011truly,langer2016robust,jodlbauer2019parallel}. In this study, our focus is specifically on the transferability of commonly used preconditioning strategies across cardiovascular FSI problems and structural formulations.

Seo et al.~\cite{Jongming1} compared iterative solvers and preconditioners for cardiovascular simulations with rigid and deformable walls, including analyses of matrix properties and parallel scalability. More recently, Sun et al.~\cite{sun2025parallel} developed a parallel solver framework for fully implicit monolithic FSI using a velocity-pressure structural formulation. These studies motivate a closer examination of how commonly used preconditioners respond to changes in structural formulation within a common monolithic framework, and whether their robustness persists across different cardiovascular FSI problems.

This study systematically investigates Krylov solvers with algebraic and physics-based preconditioners in a unified monolithic ALE framework. The numerical experiments span an idealized cylindrical benchmark and patient-specific aortic, cerebral arterial, and coronary models using both structural formulations. The principal contribution is the connection between formulation-dependent block structure and observed preconditioner robustness, supported by analysis of the pressure block and the exact and approximate Schur complements. Focused comparisons then examine how block factorization, sub-block approximations, and nested inner solves affect the cost of Schur-complement preconditioning. Together, these results identify configurations that retain convergence across the investigated cases and provide practical guidance for solver selection in cardiovascular FSI.

The remainder of the paper is organized as follows. Sections~\ref{sec:methods} and \ref{sec:linear-solvers} describe the numerical methods and preconditioning strategies. Section~\ref{sec:results} presents the solver evaluation and the analysis of Schur-complement design choices. Section~\ref{sec:discussion} discusses the formulation-dependent behavior, practical implications, and limitations of the study. Section~\ref{sec:conclusion} summarizes the principal findings.

\section{Numerical Formulation} \label{sec:methods}

We summarize the fluid and solid governing equations solved within an ALE framework for cardiovascular FSI. The fluid and structural equations are solved monolithically, while the ALE mesh-motion equation is treated separately. Linear finite elements and the generalized-$\alpha$ method are used for spatial and temporal discretization, respectively, and the resulting nonlinear system is solved by Newton iteration. Since the focus of this study is the resulting linear systems, only the aspects determining their block structure are presented here; further implementation details are provided in the supplementary material. The Krylov solvers and algebraic- and physics-based preconditioners are described separately in Section~\ref{sec:linear-solvers}. All numerical methods are implemented within the svFSI multiphysics finite element solver \citep{Zhu2022}, coupled with PETSc \citep{petsc-user-ref} to support the different linear solver configurations. This provides a consistent basis for evaluating and comparing solver performance later.

\subsection{Fluid domain}

Blood is modeled as an incompressible viscous Newtonian fluid occupying the current domain $\Omega_x^f\subset\mathbb{R}^3$. In an ALE description, the governing equations are
\begin{align}
\boldsymbol{0}
&=
\rho_0^f
\left.
\frac{\partial \boldsymbol{v}^f}{\partial t}
\right|_{\boldsymbol{x}}
+
\rho_0^f
\left(
\boldsymbol{v}^f-\hat{\boldsymbol{v}}
\right)
\cdot\nabla_x\boldsymbol{v}^f
-\nabla_x\cdot\boldsymbol{\sigma}^f \nonumber\\
& +\nabla_x p^f
-\rho_0^f\boldsymbol{b},
\\
0
&=
\nabla_x\cdot\boldsymbol{v}^f,
\label{eq:fluid}
\end{align}
where $\boldsymbol{v}^f$ and $\hat{\boldsymbol{v}}$ are the fluid and mesh velocities, respectively, $p^f$ is the pressure, $\boldsymbol{\sigma}^f$ is the viscous Cauchy stress, and $\rho_0^f$ is the fluid density.

Physiological outlet conditions are imposed through lumped-parameter models when appropriate. The flow rate of the $i$-th outlet, defined as
\begin{equation}
Q_i(t)
=
\int_{\Gamma_i^{\mathrm{out}}}
\boldsymbol{v}^f\cdot\boldsymbol{n}\,d\Gamma,
\end{equation}
is supplied to the corresponding lumped-parameter model, and the resulting pressure is returned to the three-dimensional problem as an outlet traction. Resistance, RCR, and coronary boundary conditions used in the present study are illustrated schematically with each FSI model in Section~\ref{sec:results}; details of their numerical coupling are given in Refs. \cite{esmaily2013new,moghadam2013modular}.

The fluid equations are discretized using a residual-based variational multiscale (VMS) formulation with equal-order $P_1-P_1$ interpolation of velocity and pressure \citep{bazilevs2007variational}. Of particular relevance to the linear solver, the advection term introduces asymmetry into the momentum block, while the VMS stabilization produces a nonzero pressure-pressure contribution. Therefore, the resulting linearized fluid operator differs from the classical unstabilized saddle-point form.

\subsection{Solid domain}

The vessel wall is modeled as a hyperelastic solid. Two structural formulations are considered: one uses displacement as the primary variable, and the other uses the velocity-pressure pair. For brevity, these formulations are referred to as displacement-based and velocity-based, respectively, with ``velocity-based'' used as shorthand for the velocity-pressure formulation. The two choices lead to different algebraic systems after discretization.

\subsubsection*{Displacement-based formulation}
Let $\Omega_X^s\subset\mathbb{R}^3$ denote the reference configuration and $\boldsymbol{U}^s$ the structural displacement. Introducing the structural velocity $\boldsymbol{V}^s=d\boldsymbol{U}^s/dt$, the elastodynamic equations are written as
\begin{align}
\boldsymbol{0}
&=
\frac{d\boldsymbol{U}^s}{dt}
-\boldsymbol{V}^s,
\\
\boldsymbol{0}
&=
\rho_0^s\frac{d\boldsymbol{V}^s}{dt}
-\nabla_X\cdot\boldsymbol{P}^s(\boldsymbol{U}^s)
-\rho_0^s\boldsymbol{B},
\label{eq:solid-displacement}
\end{align}
where $\rho_0^s$ is the reference density and $\boldsymbol{P}^s$ is the first Piola-Kirchhoff stress. The constitutive response is evaluated from the displacement, and pressure is not introduced as an independent structural unknown in this formulation. After coupling to the fluid system, the structural equations therefore contribute primarily to the momentum operator.

\subsubsection*{Velocity-based formulation}
In the current solid domain $\Omega_x^s\subset\mathbb{R}^3$, the Cauchy stress is decomposed as $\boldsymbol{\sigma}^s=\boldsymbol{\sigma}_{\mathrm{dev}}^s-p^s\boldsymbol{I}$. Following previous work \cite{LIU2018549}, the first-order system is
\begin{align}
\boldsymbol{0}
&=
\frac{d\boldsymbol{u}^s}{dt}
-\boldsymbol{v}^s,
\\
\boldsymbol{0}
&=
\rho^s(p^s)\frac{d\boldsymbol{v}^s}{dt}
-\nabla_x\cdot\boldsymbol{\sigma}_{\mathrm{dev}}^s(\boldsymbol{u}^s)
+\nabla_x p^s
-\rho^s(p^s)\boldsymbol{b},
\\
0
&=
\beta_\theta^s(p^s)\frac{dp^s}{dt}
+\nabla_x\cdot\boldsymbol{v}^s,
\label{eq:solid}
\end{align}
where $\boldsymbol{u}^s$, $\boldsymbol{v}^s$, and $p^s$ are the structural displacement, velocity, and pressure, respectively, and
\begin{equation}
\beta_\theta^s
=
\frac{1}{\rho^s}
\left.
\frac{\partial\rho^s}{\partial p^s}
\right|_\theta
\end{equation}
is the isothermal compressibility.

The velocity-based solid equations are discretized using a residual-based VMS formulation \citep{LIU2018549}. In contrast to the displacement-based formulation, structural pressure appears naturally as a primary variable. The momentum and mass-conservation stabilization terms also generate additional velocity-pressure coupling and a nonzero pressure-pressure contribution. This distinction is important because it changes the block structure presented to the linear solver and is examined further in Section~\ref{sec:discussion}.

\subsection{Time integration and Newton linearization}

The coupled fluid and solid equations are advanced monolithically using the generalized-$\alpha$ method for first-order systems and solved using Newton-Raphson iteration \citep{jansen2000generalized,LIU2018549}. At each Newton iteration, the discrete momentum and mass-conservation residuals are denoted by $\mathbf{R}_m$ and $\mathbf{R}_p$, respectively. Linearization, together with elimination of the solid displacement increment through the kinematic equation, gives
\begin{equation}
\begin{bmatrix}
K & G \\
D & L
\end{bmatrix}
\begin{bmatrix}
\Delta\dot{\mathbf{v}} \\
\Delta\dot{\mathbf{p}}
\end{bmatrix}
=
-
\begin{bmatrix}
\mathbf{R}_m \\
\mathbf{R}_p
\end{bmatrix},
\label{eq:newton-block-system}
\end{equation}
where $K$ is the linearized momentum block, $G$ represents the pressure contribution to the momentum equations, $D$ represents the velocity contribution to the mass-conservation equations, and $L$ contains pressure-related contributions, including those introduced by stabilization. For both structural formulations, $\Delta\dot{\mathbf{v}}$ collects the fluid and solid velocity-rate corrections after kinematic elimination.

Equation~\eqref{eq:newton-block-system} is the linear system considered throughout this study. Both structural formulations lead to the same generic $2\times2$ block structure for the entire FSI problem, but the numerical properties and sparsity patterns of the individual blocks differ because their structural unknowns and stabilization contributions are different. These formulation-dependent differences motivate the solver comparisons presented in Sections~\ref{sec:results} and \ref{sec:discussion}.

\section{Linear solvers with algebraic- and physics-based preconditioners} \label{sec:linear-solvers}

The discretized monolithic FSI formulation gives rise to large, sparse, and generally nonsymmetric linear systems, which are solved using Krylov subspace methods with different preconditioning strategies. Unless otherwise stated, GMRES is employed with a restart parameter of 200 and a maximum of 2000 iterations. This restart value follows previous cardiovascular simulation studies, where it was found to provide robust convergence behavior \citep{Jongming1}. For comparison, BiCGStab is also considered, with the maximum iteration count also set to 2000. The mesh-motion equation is solved using the conjugate gradient method with Jacobi preconditioner. Unless otherwise specified, the linear solver relative and absolute tolerances are set to $10^{-6}$ and $10^{-12}$, respectively. The Newton iteration tolerance is set to $10^{-6}$.

\subsection{Algebraic preconditioners}

We consider several algebraic preconditioning techniques used either directly on the assembled monolithic FSI system or as sub-block solvers within the physics-based preconditioners introduced in the following subsection.

The additive Schwarz method (ASM) is applied directly to the monolithic system as a parallel domain-decomposition preconditioner. The global system is partitioned into subdomains with one layer of overlap, and each local problem is approximately solved using ILU(2) factorization. Reverse Cuthill-McKee reordering is applied before factorization to reduce matrix bandwidth \citep{cuthill1969rcm}.

Algebraic multigrid methods are also considered. PETSc GAMG and Trilinos ML employ smoothed-aggregation constructions and are applied directly to the monolithic FSI system. For both implementations, a V-cycle is used, the multilevel hierarchy is constructed automatically from the assembled matrix, and the number of levels is determined adaptively according to the problem size. The same parameter settings are retained across the investigated cases to provide a consistent basis for comparison.

BoomerAMG from the hypre package \citep{falgout2002hypre} represents a classical AMG approach and is used differently in the present study. Although BoomerAMG was also tested as a preconditioner for the full monolithic FSI system, it showed substantially poorer performance than GAMG and ML. Therefore, it is not included in the monolithic solver comparisons. Instead, BoomerAMG is used within the physics-based block preconditioners in Section \ref{sec:physics-based-preconditioners} to approximate selected sub-block inverse actions. BoomerAMG is configured with HMIS coarsening, extended+i interpolation, and hybrid Gauss-Seidel relaxation, following settings used in previous cardiovascular flow studies \citep{baker2011multigrid,LIU-nest}.

\subsection{Physics-based preconditioners}\label{sec:physics-based-preconditioners}

We consider physics-based preconditioners that exploit the block structure of the linear system in Eq.~\eqref{eq:block-mat}. Two classes are investigated: simple block splittings based on the original diagonal blocks $K$ and $L$, and Schur-complement preconditioners that incorporate the strong coupling between the momentum and continuity equations.

\subsubsection{Simple block splittings}
The simplest physics-based approaches are additive and multiplicative block preconditioners. The additive preconditioner is of block Jacobi type and neglects both off-diagonal coupling blocks, essentially decoupling the velocity and pressure subproblems during preconditioning.
\begin{equation}
\mathcal{P}_{\mathrm{add}}
=
\begin{bmatrix}
K & 0 \\
0 & L
\end{bmatrix}.
\label{eq:block-add}
\end{equation}
In contrast, the multiplicative preconditioner is of block Gauss-Seidel type and retains one direction of the coupling,
\begin{equation}
\mathcal{P}_{\mathrm{mult}}
=
\begin{bmatrix}
K & 0 \\
D & L
\end{bmatrix}.
\label{eq:block-mult}
\end{equation}
In both cases, the inverse actions of $K$ and $L$ are approximated using BoomerAMG.

\subsubsection{Schur-complement preconditioners}
Although simple block splittings are inexpensive, using $L$ alone provides only a limited approximation of the coupled pressure dynamics. For strongly coupled FSI systems, this motivates the use of a Schur-complement reduction that incorporates the coupling through $G$ and $D$ \citep{Benzi_Golub_Liesen_2005}. The exact block factorization of the monolithic system matrix is
\begin{equation}
A
=
\begin{bmatrix}
I & 0 \\
DK^{-1} & I
\end{bmatrix}
\begin{bmatrix}
K & 0 \\
0 & S
\end{bmatrix}
\begin{bmatrix}
I & K^{-1}G \\
0 & I
\end{bmatrix},
\label{A:factorize}
\end{equation}
where
\begin{equation}
S=L-DK^{-1}G
\label{eq:exact-schur}
\end{equation}
is the exact Schur complement. Compared with $L$, the operator $S$ incorporates the influence of the momentum variables through the coupling term $DK^{-1}G$.

The exact Schur complement is generally unsuitable for explicit construction because $K^{-1}$ can be dense. We therefore use the sparse approximation
\begin{equation}
\tilde{S}
=
L-D\bigl(\mathrm{diag}(K)\bigr)^{-1}G.
\label{eq:approx-schur}
\end{equation}
The block factorization in Eq.~\eqref{A:factorize} and the approximation in Eq.~\eqref{eq:approx-schur} control two distinct aspects of the preconditioner. The former helps to design how much of the velocity-pressure coupling is retained, whereas the latter determines how the reduced pressure operator is approximated.

Four block structures are considered for preconditioning, as summarized in Table~\ref{tab:factorization}. The diagonal form omits both coupling factors, the lower and upper triangular forms retain one coupling direction, and the full form retains both, thereby most closely reproducing the exact block factorization of $A$. In all four cases, the exact Schur complement $S$ is replaced by the sparse approximation $\tilde{S}$ defined in Eq.~\eqref{eq:approx-schur}. The inverse actions of $K$ and $\tilde{S}$ are approximated using BoomerAMG.

\begin{table}[htbp]
\centering
\small
\caption{Block structures considered for Schur-complement preconditioning. The exact Schur complement $S$ is replaced by the sparse approximation $\tilde{S}$ defined in Eq.~\eqref{eq:approx-schur}.}
\label{tab:factorization}
\renewcommand{\arraystretch}{1.4}
\begin{tabular}{ll}
\toprule
\textbf{Block structure} & \textbf{Preconditioner} \\
\midrule

Diagonal &
$
\displaystyle
\mathcal{P}_{\mathrm{diag}}
=
\begin{bmatrix}
K & 0 \\
0 & \tilde{S}
\end{bmatrix}
$
\\[2.5ex]

Lower triangular &
$
\displaystyle
\mathcal{P}_{\mathrm{lower}}
=
\begin{bmatrix}
I & 0 \\
DK^{-1} & I
\end{bmatrix}
\begin{bmatrix}
K & 0 \\
0 & \tilde{S}
\end{bmatrix}
$
\\[3.5ex]

Upper triangular &
$
\displaystyle
\mathcal{P}_{\mathrm{upper}}
=
\begin{bmatrix}
K & 0 \\
0 & \tilde{S}
\end{bmatrix}
\begin{bmatrix}
I & K^{-1}G \\
0 & I
\end{bmatrix}
$
\\[3.5ex]

Full &
$
\displaystyle
\mathcal{P}_{\mathrm{full}}
=
\begin{bmatrix}
I & 0 \\
DK^{-1} & I
\end{bmatrix}
\begin{bmatrix}
K & 0 \\
0 & \tilde{S}
\end{bmatrix}
\begin{bmatrix}
I & K^{-1}G \\
0 & I
\end{bmatrix}
$
\\
\bottomrule
\end{tabular}
\end{table}

We additionally consider the nested block preconditioning strategy proposed in Ref. \cite{LIU-nest} for the incompressible Navier–Stokes equations and apply it here to the monolithic FSI system. The key difference from the block preconditioners described above, which are of non-nested type, is the treatment of the required block inverse actions. Rather than using a single algebraic preconditioner application to approximate a block inverse, the nested approach obtains these actions through inner GMRES solves. Hereafter, GMRES$(N,\varepsilon)$ denotes GMRES with a maximum of $N$ iterations and relative tolerance $\varepsilon$.

The nested preconditioner considered here uses a lower triangular block factorization of the monolithic system, with inner solves following the strategy in Ref. \cite{LIU-nest}. The momentum-block inverse actions are approximated by GMRES$(N_0,\varepsilon_0)$ preconditioned by BoomerAMG, while the Schur-complement system is solved by GMRES$(N_1,\varepsilon_1)$ with a BoomerAMG preconditioner constructed from $\tilde{S}$. The exact Schur
complement $S$ is not formed explicitly. Instead, its action on a vector $\boldsymbol{v}$ is evaluated as
\begin{equation}
S\boldsymbol{v}=L\boldsymbol{v}-DK^{-1}G\boldsymbol{v},
\label{eq:schur-action}
\end{equation}
where the additional action of $K^{-1}$ required in Eq.~\eqref{eq:schur-action} is approximated by GMRES$(N_0,\varepsilon_2)$, again preconditioned by BoomerAMG. Thus, the nested strategy introduces an additional Krylov level within the Schur-complement evaluation. Because the resulting preconditioning operation varies with the convergence of the inner solves, flexible GMRES (FGMRES) is used as the outer Krylov solver \citep{saad1993flexible}.

For most solver comparisons in Section~\ref{sec:results}, we use the non-nested lower triangular Schur-complement preconditioner as the default Schur-complement configuration, hereafter denoted Block-Schur. The rationale for this choice is presented in Section~\ref{sec:sub-block-discussion}, where we examine the effects of block factorization and sub-block approximation and compare the non-nested approach with the nested strategy described above.

\section{Results} \label{sec:results}

All computations are performed on a high-performance computing cluster. Each compute node is equipped with Intel Xeon Gold 6258R processors with 56 cores and 192 GB of memory. Inter-node communication is supported by a 100 Gbps InfiniBand network. All cases are discretized using linear tetrahedral elements and solved in parallel using MPI. The corresponding mesh and parallelization details are summarized in Table \ref{tab:patient_mesh}. The number of MPI processes is selected to maintain a workload of approximately $25{,}000$ elements per core, following the parallel efficiency study in Ref. \cite{Jongming1}, which also employed svFSI. The solver modifications do not fundamentally alter the code structure of svFSI, and the parallel performance remains similar to that reported in Ref. \cite{Jongming1}. A dedicated parallel-efficiency study is provided in the supplementary material.

\begin{table}[htbp]
\centering
\caption{Mesh parameters and computational resources for all cases.}
\label{tab:patient_mesh}
\begin{tabular}{lccccc}
\toprule
\textbf{Case} & \multicolumn{2}{c}{\textbf{Lumen}} & \multicolumn{2}{c}{\textbf{Wall}} & \textbf{FSI} \\
\cmidrule(lr){2-3} \cmidrule(lr){4-5} \cmidrule(lr){6-6}
 & Points & Elements & Points & Elements & Cores \\
\midrule
Cylindrical pipe & 166,899 & 1,002,436 & 50,909 & 192,668 & 48 \\
Aorta & 430,798 & 2,569,521 & 159,284 & 714,474 & 132 \\
Cerebral artery & 314,479 & 1,635,612 & 319,725 & 1,276,350 & 112 \\
Coronary artery & 444,323 & 2,625,613 & 181,812 & 723,132 & 132 \\
\bottomrule
\end{tabular}
\end{table}

\subsection{Cylindrical pipe benchmark} \label{sec:sub-pipe}

\begin{figure}[htbp]
    \centering
    \includegraphics[width=0.8\linewidth]{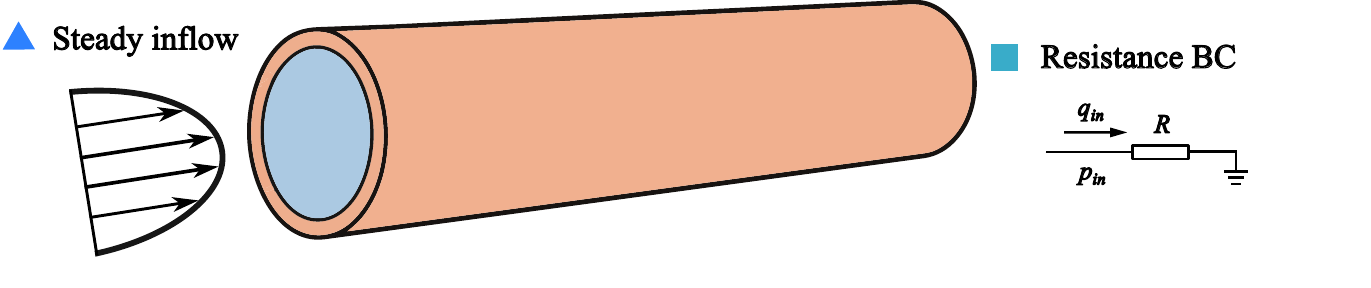}
    \caption{The cylindrical pipe model with a resistance boundary condition.}
    \label{fig:pipe-model}
\end{figure}

We first consider a classical cylindrical pipe benchmark that approximates the ascending aorta \citep{Jongming1}. As shown in Figure \ref{fig:pipe-model}, the model consists of a straight vessel with a radius of $2$ cm, a length of $30$ cm, and a wall thickness of $0.2$ cm. A steady parabolic velocity profile corresponding to a volumetric flow rate of $83$ mL/s is prescribed at the inlet. At the outlet, a resistance boundary condition with $R=1600 \mathrm{g\cdot cm^{-4}\cdot s^{-1}}$ is imposed. The vessel wall is modeled as a Neo-Hookean material with a Poisson ratio of $0.3$ and a Young’s modulus of 1.0 MPa. The simulations are performed over 30 time steps with a time step size of 1 ms. The reported wall time represents the average of two independent runs.

\begin{figure}[htbp]
    \centering
    \includegraphics[width=0.9\linewidth]{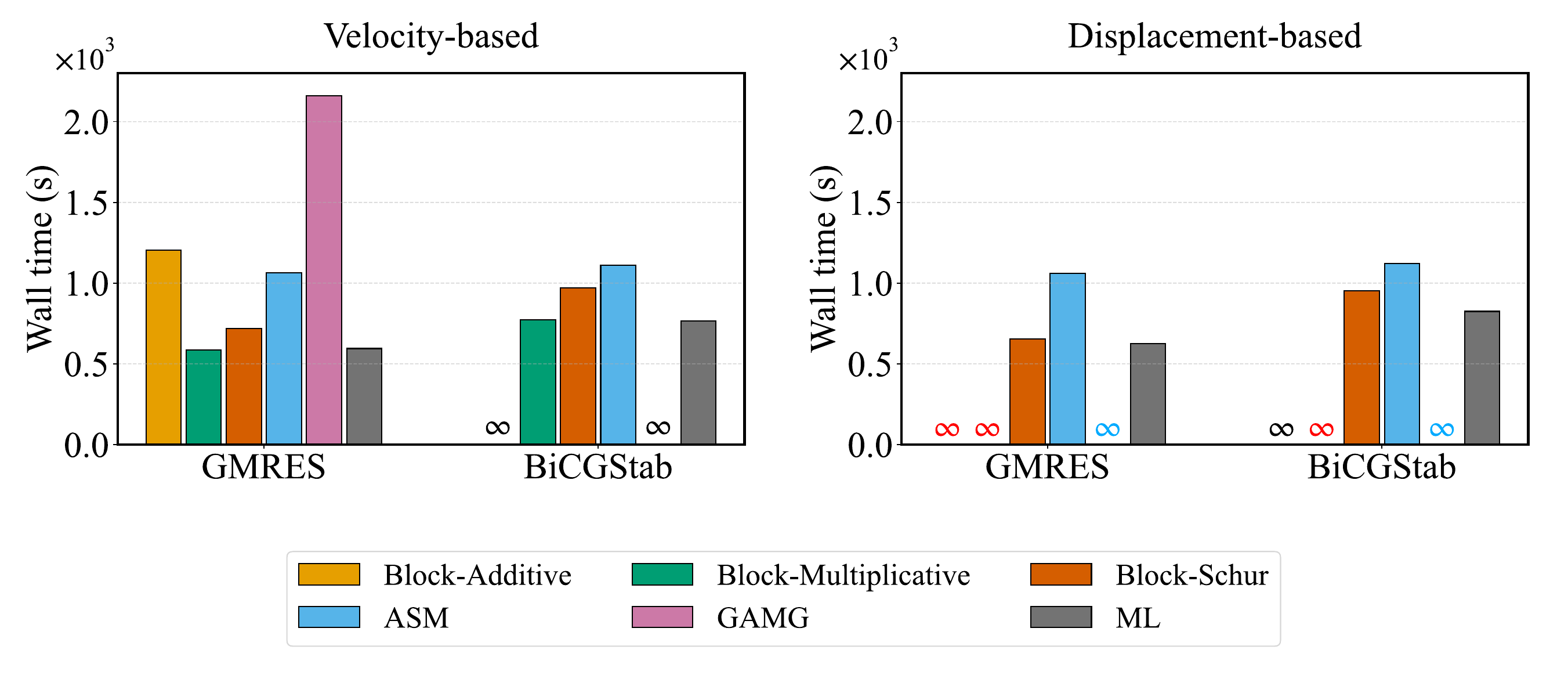}
    \caption{Wall time of the investigated solver-preconditioner combinations for the cylindrical pipe benchmark using the velocity-based (left) and displacement-based (right) structural formulations. $\infty$ indicates non-convergence, with colors distinguishing the failure modes: black, Newton solver failure; red, linear solver breakdown; blue, maximum linear solver iterations reached.}
    \label{fig:pipe}
\end{figure}

\begin{table}[htbp]
\centering
\small
\caption{Accumulated Krylov iteration counts and fraction of total wall time spent solving the monolithic FSI system for the cylindrical pipe benchmark. $N_{\mathrm{GMRES}}$ and $N_{\mathrm{BiCGStab}}$ denote the accumulated iteration counts.
}
\label{tab:pipe_case_table}
\begin{tabular}{lcccc}
\toprule
\textbf{Case 1: Cylindrical Pipe} & \multicolumn{2}{c}{\textbf{GMRES}} & \multicolumn{2}{c}{\textbf{BiCGStab}} \\
\cmidrule(r){2-3} \cmidrule(l){4-5}
\textbf{PC} & $N_{\text{GMRES}}$ & $t_{\text{FSI}}/t_{\text{Total}}$   & $N_{\text{BiCGStab}}$ & $t_{\text{FSI}}/t_{\text{Total}}$   \\
\midrule
\multicolumn{5}{l}{\textbf{Velocity-based structural formulation}} \\
\midrule
Block-Additive       & $2.24\times10^4$ & $93.86\%$ & N/A & N/A \\
Block-Multiplicative & $8.83\times10^3$ & $89.79\%$ & $7.17\times10^3$ & $92.79\%$ \\
Block-Schur          & $7.16\times10^3$ & $91.56\%$ & $6.45\times10^3$ & $94.35\%$ \\
ASM                  & $4.23\times10^3$ & $94.92\%$ & $2.59\times10^3$ & $94.07\%$ \\
GAMG                 & $3.14\times10^4$ & $97.06\%$ & N/A & N/A  \\
ML                   & $5.78\times10^3$ & $89.37\%$ & $4.03\times10^3$ & $92.67\%$ \\
\midrule
\addlinespace[1.5ex]

\multicolumn{5}{l}{\textbf{Displacement-based structural formulation}} \\
\midrule
Block-Additive       & N/A & N/A  & N/A & N/A  \\
Block-Multiplicative & N/A & N/A  & N/A & N/A  \\
Block-Schur          & $6.27\times10^3$ & $91.34\%$ & $6.49\times10^3$ & $94.04\%$ \\
ASM                  & $4.33\times10^3$ & $94.79\%$ & $2.67\times10^3$ & $94.04\%$ \\
GAMG                 & N/A & N/A  & N/A & N/A  \\
ML                   & $6.03\times10^3$ & $89.80\%$ & $5.31\times10^3$ & $92.73\%$ \\
\bottomrule
\end{tabular}
\end{table}

Figure~\ref{fig:pipe} compares the performance of the investigated preconditioners for the two structural formulations. The physics-based methods include the additive and multiplicative block preconditioners, denoted Block-Additive and Block-Multiplicative, respectively, together with the reference non-nested lower triangular Schur-complement preconditioner, denoted Block-Schur. The monolithic algebraic preconditioners are ASM, GAMG, and ML.

For the velocity-based formulation, all configurations converge except Block-Additive and GAMG when combined with BiCGStab. Among the physics-based preconditioners, Block-Multiplicative gives the shortest wall time. Block-Schur also converges with both Krylov solvers but requires a longer wall time. Block-Additive is less efficient and fails with BiCGStab. Among the monolithic algebraic preconditioners, ML gives the shortest wall time, ASM shows intermediate performance, and GAMG is the most expensive with GMRES.

For the displacement-based formulation, solver robustness deteriorates significantly. Block-Additive, Block-Multiplicative, and GAMG fail with both GMRES and BiCGStab, whereas Block-Schur, ASM, and ML remain convergent. Among the convergent methods, ML gives the shortest wall time, while Block-Schur also maintains competitive performance. The wall times of methods that converge under both formulations remain of the same order; the dominant effect of the formulation change is therefore the loss of convergence for several preconditioners rather than a systematic effect on computational cost. These results show that the structural formulation can substantially alter preconditioner robustness even for the same physical benchmark.

Table~\ref{tab:pipe_case_table} reports the accumulated Krylov iteration counts and the fraction of total runtime spent in the FSI solve. For all convergent configurations, the linear solver accounts for approximately 90-97\% of the total wall time, confirming that it dominates the computational cost. The patient-specific cases studied later show the same behavior. The formulation change does not systematically increase the iteration counts of the methods that remain convergent. Its principal effect is instead the loss of convergence of Block-Additive, Block-Multiplicative, and GAMG. Although ML performs the best for this benchmark, it does not persist uniformly in later patient-specific cases.

\subsection{Patient specific cases} \label{sub:patient-cases}

While the cylindrical pipe benchmark provides a controlled setting for solver evaluation, it does not capture the geometric complexity, transient flow, and physiological boundary conditions encountered in patient-specific FSI simulations. We further consider three image-based cases selected to represent distinct computational challenges: the aorta, characterized by large-vessel, advection-dominated flow; the cerebral arterial network, featuring complex vascular connectivity, loops, and multiple outlets; and the coronary artery, which requires sophisticated boundary conditions to account for the effects of myocardial contraction \citep{taylor2013}.

The mesh sizes and computational resources used for each model are summarized in Table \ref{tab:patient_mesh}. The aortic and coronary cases have similar mesh sizes and fluid-to-solid element ratios, whereas the cerebral case has a more balanced distribution of fluid and solid elements. For all simulations here, the vessel wall is modeled as a Neo-Hookean material with a Poisson ratio of $0.3$ and a Young’s modulus of 0.685 MPa. Each case is run from the beginning of the cardiac cycle to peak systole, and the reported results are averaged over the two runs.

\subsubsection{Aorta}

\begin{figure}[htbp]
    \centering
    \includegraphics[width=0.45\linewidth]{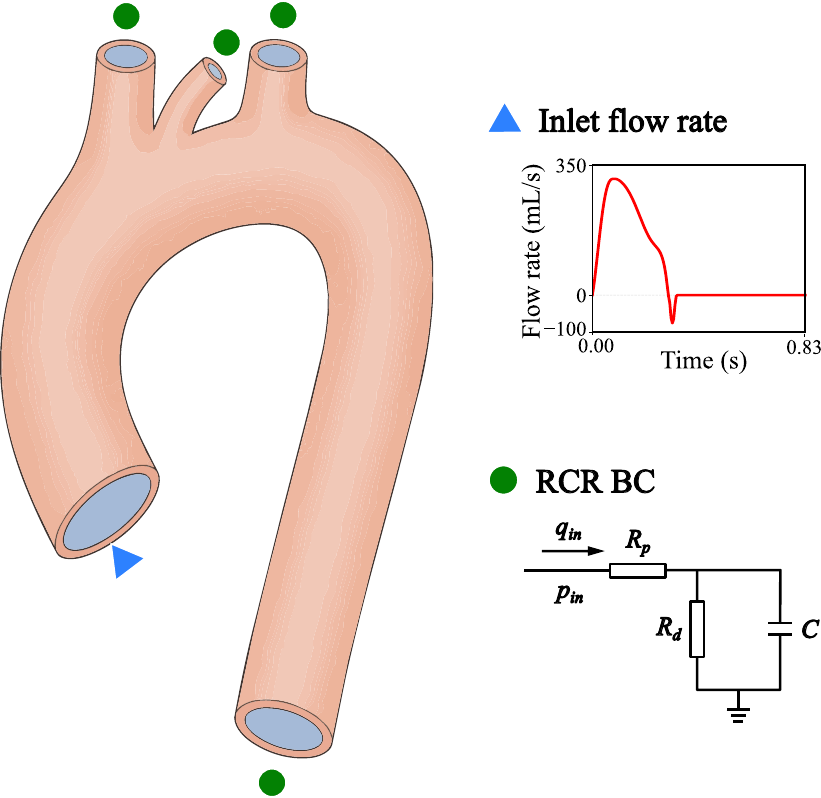}
    \caption{The aorta model with RCR boundary conditions.}
    \label{fig:aneurysm-model}
\end{figure}

\begin{figure}[htbp]
    \centering
    \includegraphics[width=0.9\linewidth]{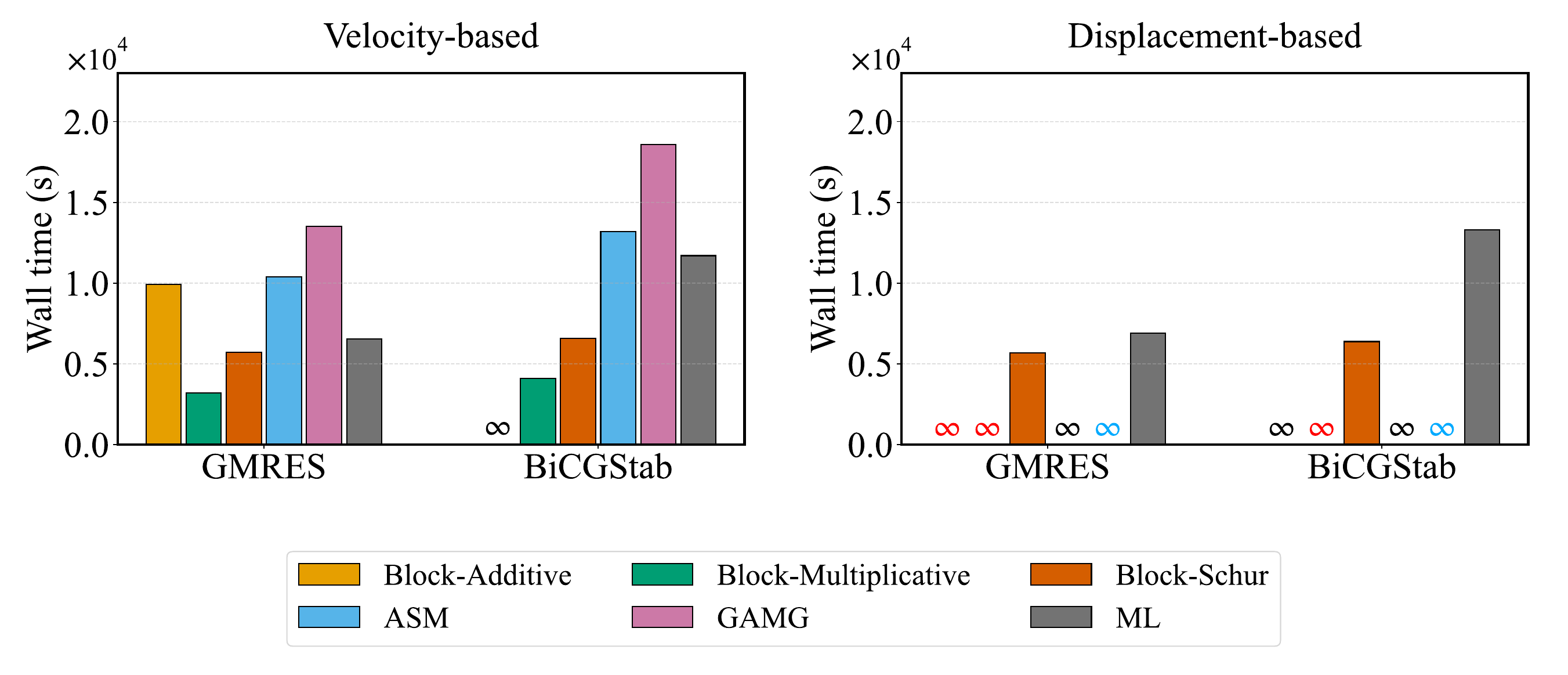}
    \caption{Wall time of the investigated solver-preconditioner combinations for the aorta case using the velocity-based (left) and displacement-based (right) structural formulations. $\infty$ indicates non-convergence, with colors distinguishing the failure modes: black, Newton solver failure; red, linear solver breakdown; blue, maximum linear solver iterations reached.}
    \label{fig:aneurysm}
\end{figure}

The aortic simulation is performed using a time-step size of $1$ ms, with $200$ time steps. A pulsatile parabolic velocity profile is prescribed at the inlet, and all outlets are coupled to RCR boundary conditions, as shown in Figure \ref{fig:aneurysm-model}. The peak inlet Reynolds number is around $3800$.

The solver performance for this case is shown in Figure \ref{fig:aneurysm}. For the velocity-based formulation, all preconditioners except Block-Additive with BiCGStab converge successfully. Among the physics-based preconditioners, Block-Multiplicative achieves the shortest solution time, followed by Block-Schur. The monolithic algebraic preconditioners require considerably longer computation, with ASM, GAMG, and ML exhibiting wall times approximately two to four times greater than those of the best-performing physics-based preconditioners.

The displacement-based formulation leads to a dramatic reduction in solver robustness. Only Block-Schur and ML converge successfully, whereas the rest fails to converge with either GMRES or BiCGStab. Among the convergent methods, Block-Schur is consistently faster than ML for both Krylov solvers. Again, the wall times of the convergent methods are similar to those obtained with the velocity-based formulation.

Compared with the cylindrical benchmark, the patient-specific aortic simulation substantially increases the disparity in both robustness and computational efficiency among the investigated preconditioners.

\subsubsection{Cerebral artery}

\begin{figure}[htbp]
    \centering
    \includegraphics[width=0.4\linewidth]{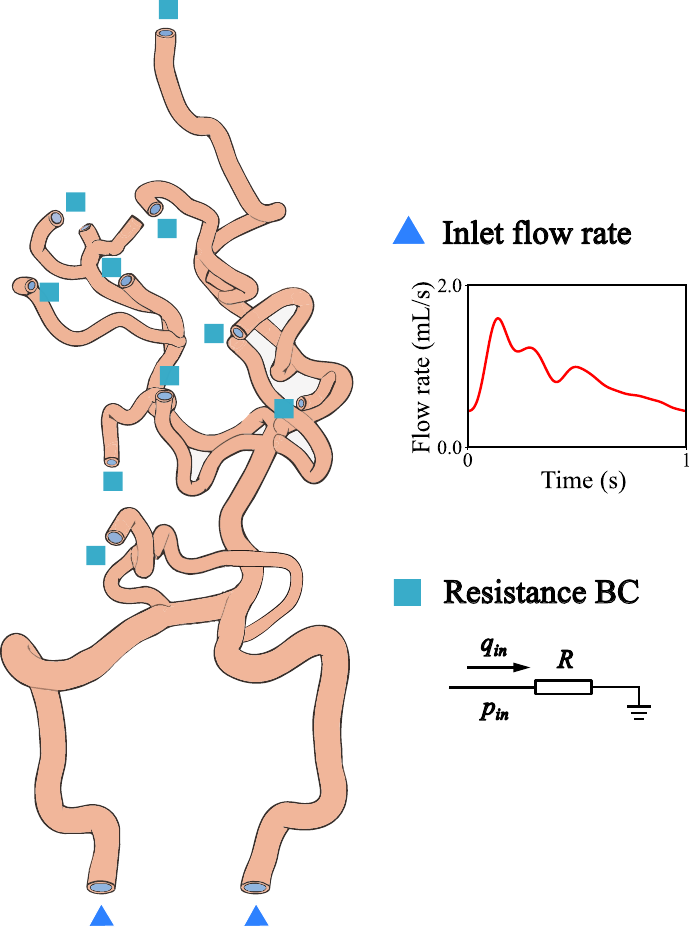}
    \caption{The cerebral artery model with resistance boundary conditions.}
    \label{fig:cerebral-model}
\end{figure}

\begin{figure}[htbp]
    \centering
    \includegraphics[width=0.9\linewidth]{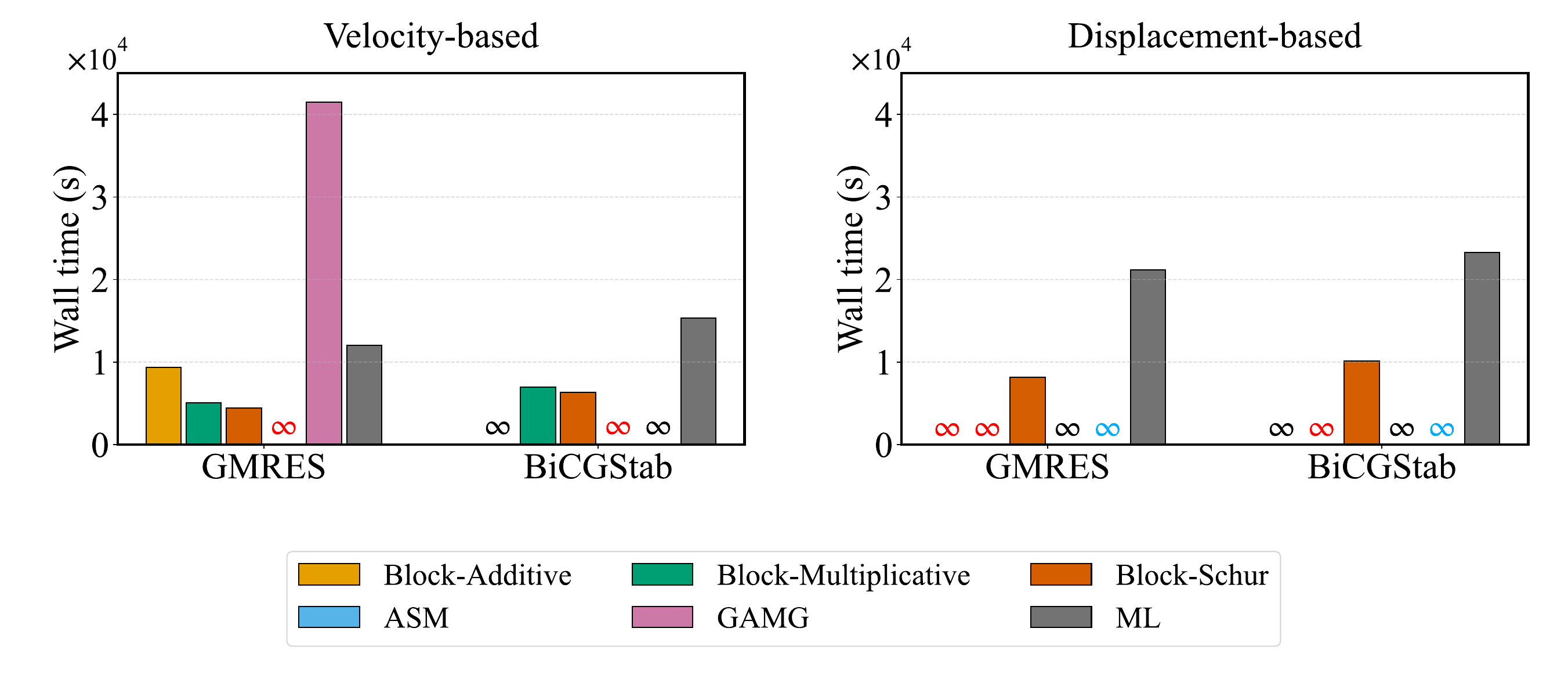}
    \caption{Wall time of the investigated solver-preconditioner combinations for the cerebral artery case using the velocity-based (left) and displacement-based (right) structural formulations. $\infty$ indicates non-convergence, with colors distinguishing the failure modes: black, Newton solver failure; red, linear solver breakdown; blue, maximum linear solver iterations reached.}
    \label{fig:cerebral}
\end{figure}

The cerebral simulation is performed with a time step size of 5 ms over 100 time steps, reaching peak systole. A pulsatile parabolic velocity profile is prescribed at both inlets, while all outlets are coupled with resistance boundary conditions (Figure \ref{fig:cerebral-model}). The peak Reynolds number at the inlet is approximately 110.

The solver performance is summarized in Figure \ref{fig:cerebral}. For the velocity-based formulation, Block-Multiplicative and Block-Schur achieve the shortest solution times among the convergent methods. Block-Additive also converges with GMRES, although at approximately twice the computational cost, whereas it fails with BiCGStab. Among the monolithic algebraic preconditioners, GAMG converges only with GMRES and exhibits the longest wall time, ASM fails to converge with either Krylov solver, and ML converges with both GMRES and BiCGStab but requires substantially longer wall time than the best-performing physics-based preconditioners.

For the displacement-based formulation, the convergence behavior is similar to that observed in the aortic simulation: only Block-Schur and ML converge with both GMRES and BiCGStab. Among these methods, Block-Schur consistently outperforms ML, reducing the wall time by more than a factor of two. In contrast to the aortic case, however, the convergent methods require noticeably longer wall times with the displacement-based formulation than with the velocity-based formulation.

\subsubsection{Coronary artery}

\begin{figure}[htbp]
    \centering
    \includegraphics[width=0.55\linewidth]{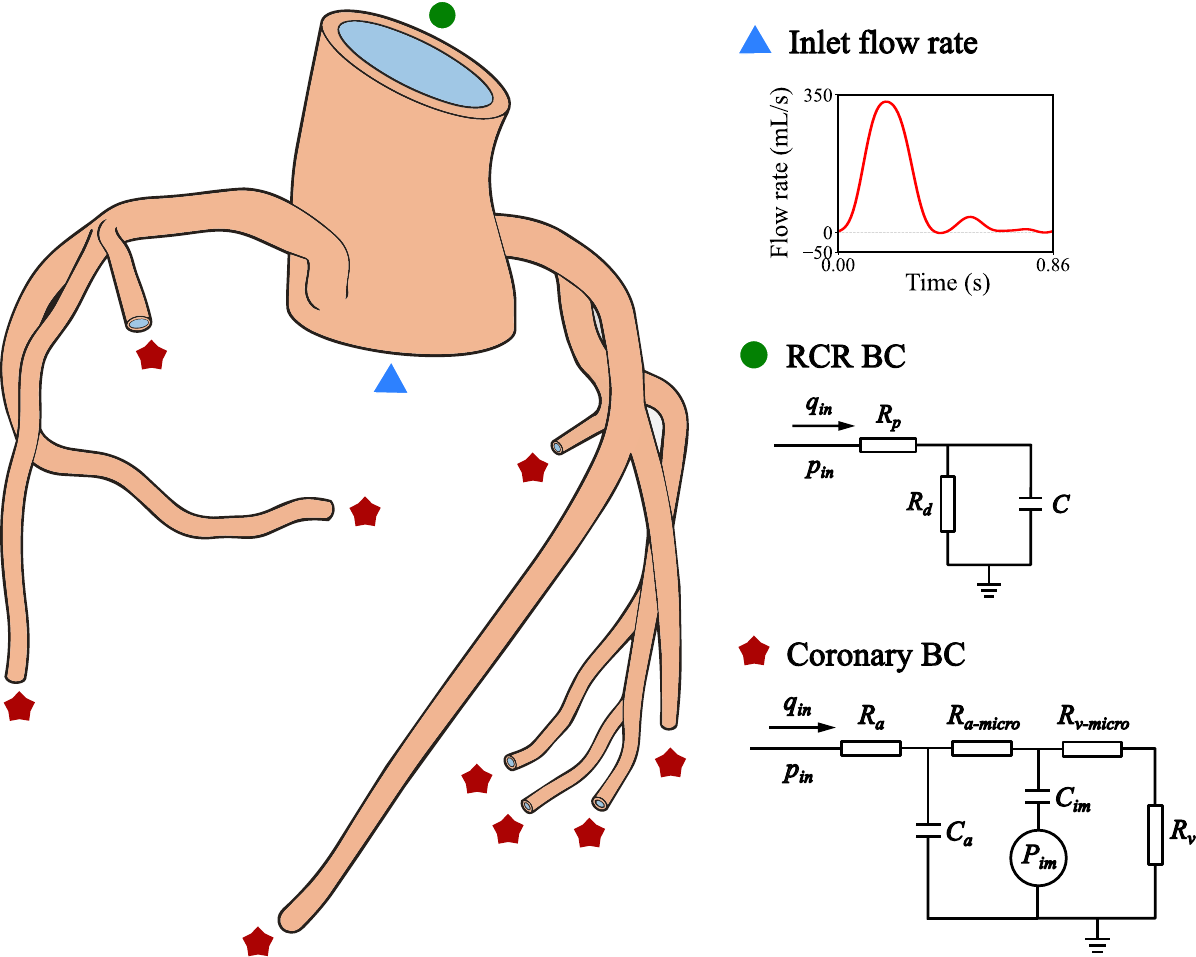}
    \caption{The coronary artery model with the RCR boundary condition at the aorta outlet and the coronary boundary conditions at side branches.}
    \label{fig:coronary-model}
\end{figure}

\begin{figure}[htbp]
    \centering
    \includegraphics[width=0.9\linewidth]{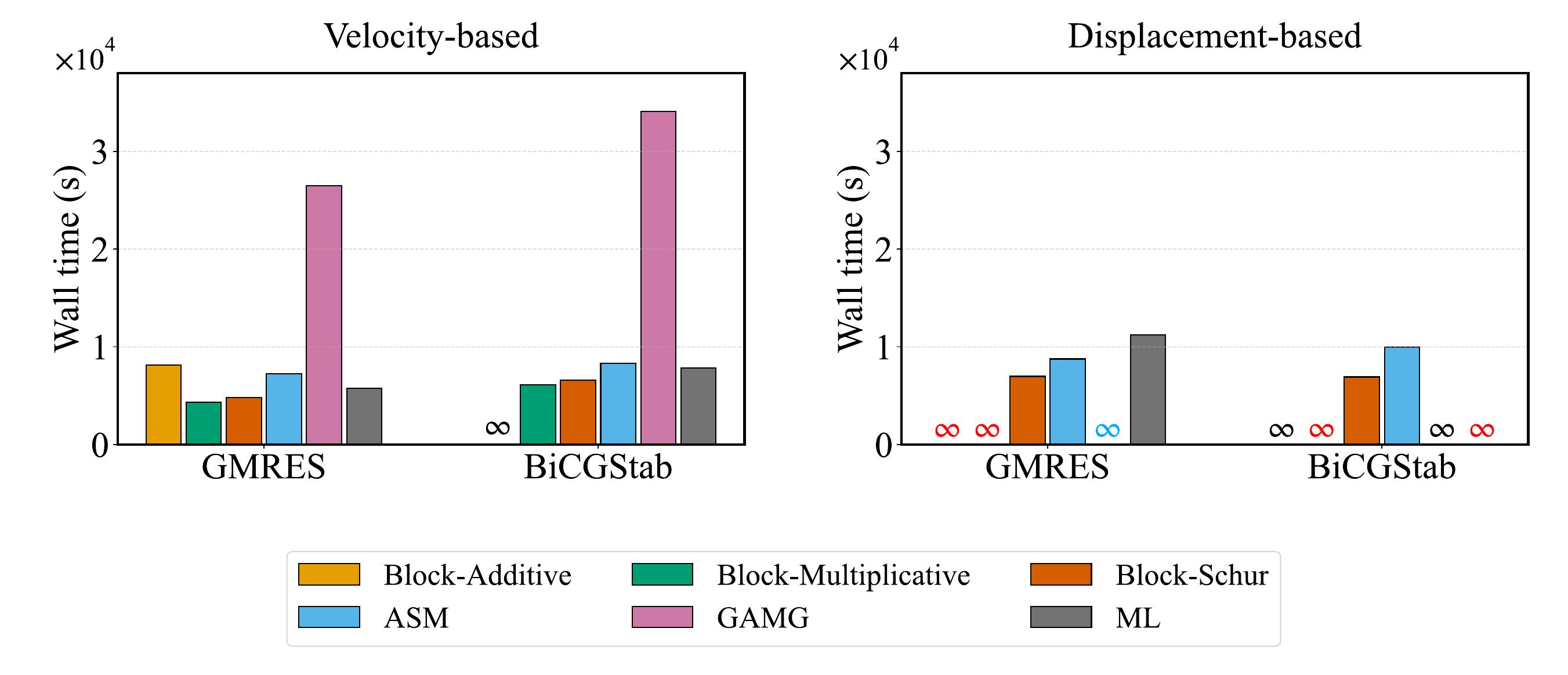}
    \caption{Wall time of the investigated solver-preconditioner combinations for the coronary artery case using the velocity-based (left) and displacement-based (right) structural formulations. $\infty$ indicates non-convergence, with colors distinguishing the failure modes: black, Newton solver failure; red, linear solver breakdown; blue, maximum linear solver iterations reached.}
    \label{fig:coronary}
\end{figure}

The coronary simulation is performed with a time step size of 1 ms over 150 time steps. As shown in Figure \ref{fig:coronary-model}, a physiological inflow waveform is prescribed at the inlet, while the RCR boundary condition is applied at the aortic outlet and coronary boundary conditions are applied at the coronary outlets \citep{taylor2013}. The peak Reynolds number at the left coronary branch is about 100.

The solver performance is summarized in Figure \ref{fig:coronary}. For the velocity-based formulation, all preconditioners except Block-Additive with BiCGStab converge successfully. Among the physics-based preconditioners, Block-Multiplicative achieves the shortest wall time, closely followed by Block-Schur. Among the monolithic algebraic preconditioners, ASM and ML require slightly longer wall times, whereas GAMG is substantially slower than all other convergent methods.

For the displacement-based formulation, the convergence behavior again changes significantly. Only Block-Schur and ASM converge with both GMRES and BiCGStab, while ML converges only with GMRES. Block-Additive, Block-Multiplicative, and GAMG fail to converge with either Krylov solver. Among the convergent methods, Block-Schur achieves the shortest wall time for both Krylov solvers. All convergent methods require slightly longer wall times than those obtained with the velocity-based formulation.

\subsection{Design choices in Schur-complement block preconditioning} \label{sec:sub-block-discussion}

The previous results show that Schur-complement preconditioning provides the most consistent convergence behavior across all simulations. We next investigate how its principal design choices affect computational cost. The aortic model at peak systole is used as the test case, with GMRES as the Krylov solver. The reported wall time represents the average of two independent runs of a single linear solve.

\subsubsection{Impact of factorization types}

We first investigate the influence of block factorization on the performance of the Schur-complement preconditioner. Table \ref{tab:factorization} summarizes the four factorization strategies considered. The diagonal factorization neglects both off-diagonal blocks and therefore completely decouples the two diagonal subproblems during preconditioning. The lower and upper triangular factorizations each retain one off-diagonal coupling block, whereas the full block factorization preserves both coupling blocks and provides the closest approximation to the exact block factorization, i.e. Equation \eqref{A:factorize}. For all four variants, both $K$ and $\tilde{S}$ are approximated using BoomerAMG so that the effect of the block factorization can be isolated.

\begin{figure}[htbp]
    \centering
    \includegraphics[width=0.9\linewidth]{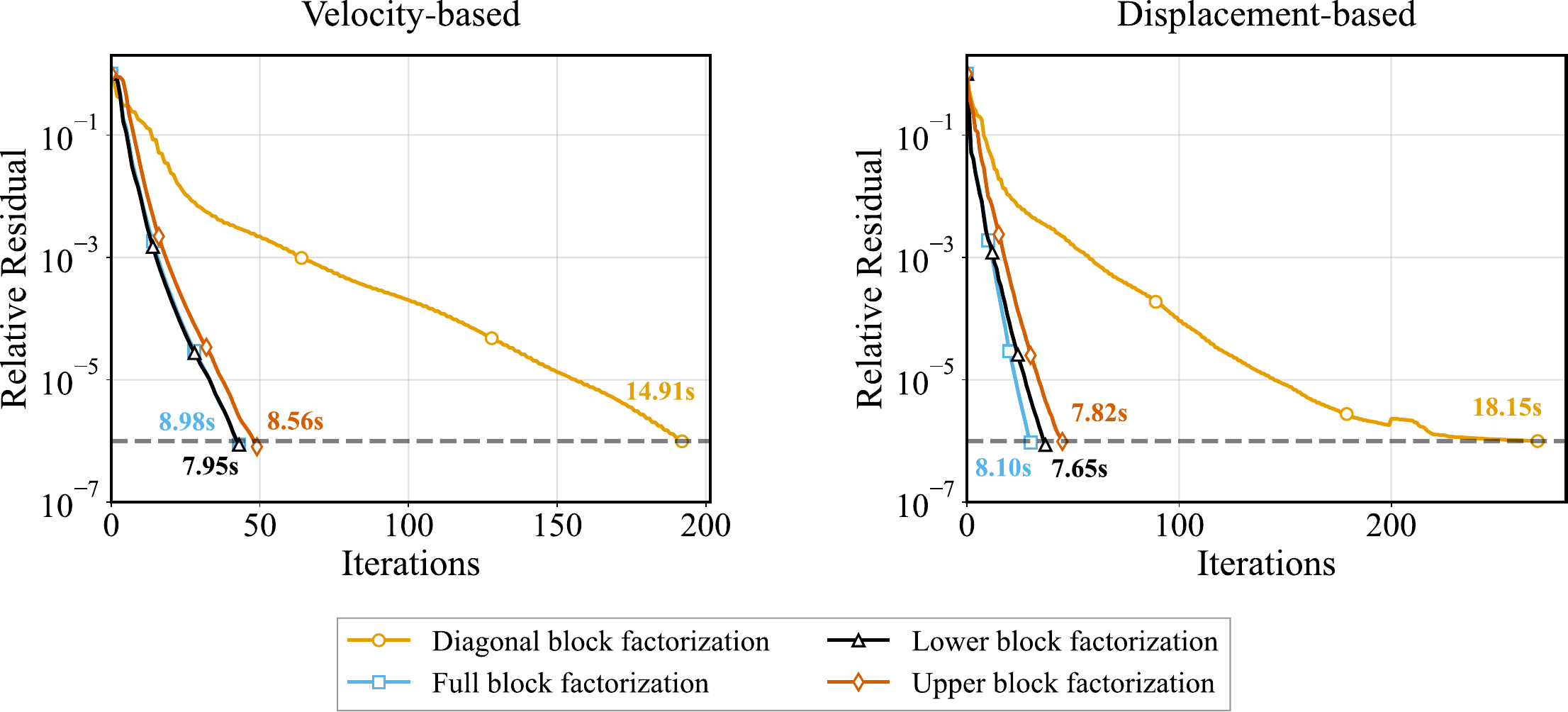}
    \caption{Effect of block factorization on the iteration count and wall time of the Schur-complement preconditioner, using the velocity-based (left) and displacement-based (right) structural formulations.}
    \label{fig:convergence-factorization}
\end{figure}

Figure \ref{fig:convergence-factorization} compares the four block factorization strategies for the velocity- and displacement-based formulations. The same trend is observed in both cases. The diagonal factorization converges substantially more slowly than the other variants, requiring approximately four to five times more Krylov iterations and nearly twice the wall time to reach the prescribed tolerance. The lower triangular, upper triangular, and full factorizations exhibit similar convergence behavior, reducing the residual by more than six orders of magnitude within approximately 50 iterations. Although the full factorization requires the fewest iterations, the lower triangular factorization consistently yields the shortest wall time, followed closely by the upper triangular factorization. Overall, the lower triangular factorization provides the best balance between convergence rate and computational cost and is therefore used in previous comparisons.

\subsubsection{Effect of sub-block approximations}

\begin{figure}[htbp]
    \centering
    \includegraphics[width=0.9\linewidth]{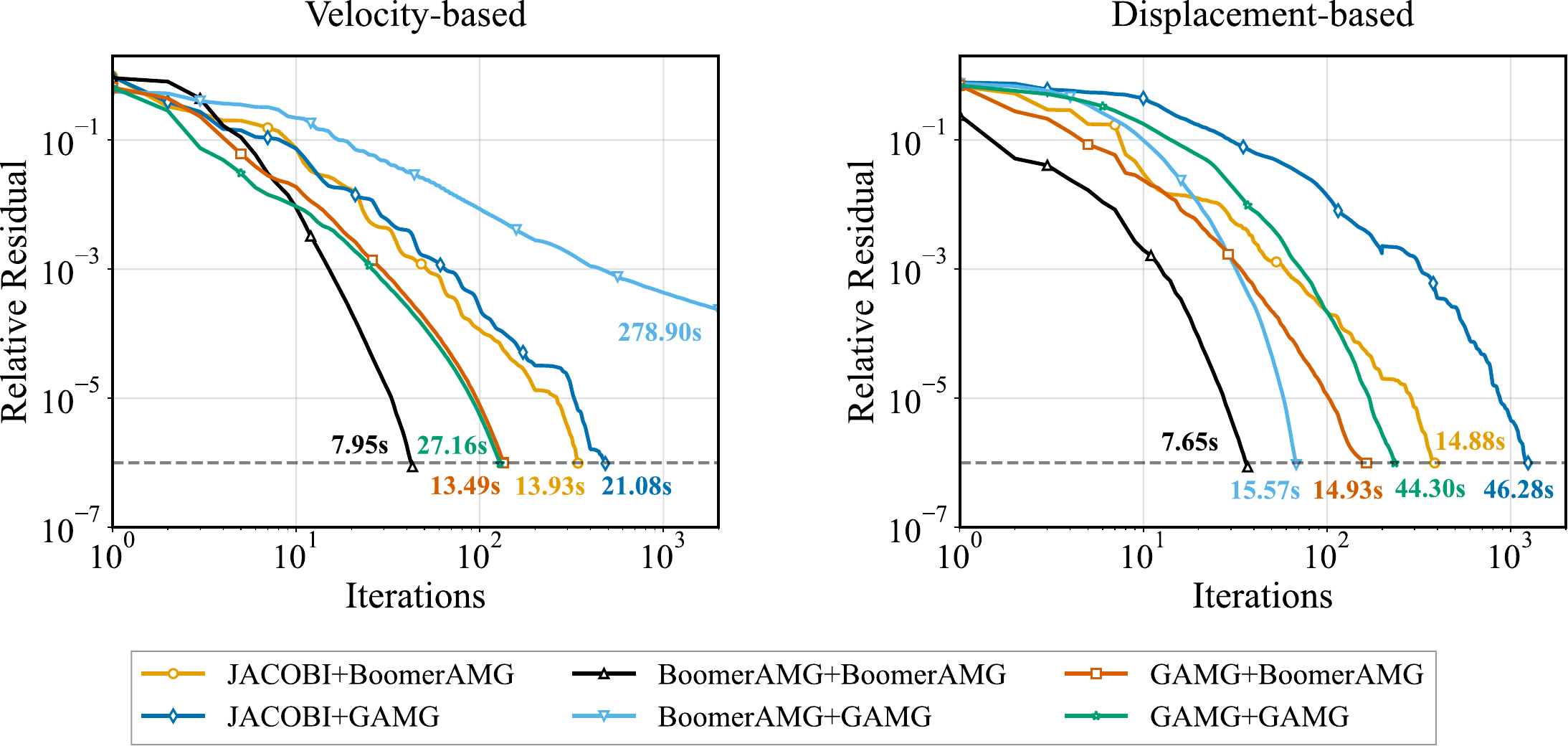}
    \caption{Effect of sub-block approximations on the iteration count and wall time of the Schur-complement preconditioner, using the velocity-based (left) and displacement-based (right) structural formulations. For each configuration, the first and second methods denote the approximations used for $K^{-1}$ and $\tilde{S}^{-1}$, respectively.}
    \label{fig:convergence-block-approx}
\end{figure}

We next examine the influence of the approximations used for the velocity block $K$ and the Schur-complement block $\tilde{S}$. BoomerAMG and GAMG are considered for both blocks, while Jacobi is additionally tested for $K$ to assess the trade-off between approximation quality and per-iteration cost. In the notation below, the first and second methods denote the approximations applied to $K$ and $\tilde{S}$, respectively.

Figure~\ref{fig:convergence-block-approx} shows that the performance of Block-Schur depends on the combined choice of sub-block approximations rather than on either approximation independently. For both structural formulations, BoomerAMG+BoomerAMG provides the shortest wall time, at $7.95$ s for the velocity-based formulation and $7.65$ s for the displacement-based formulation. Keeping BoomerAMG for the $\tilde{S}^{-1}$ action, replacing BoomerAMG by either Jacobi or GAMG for the $K^{-1}$ action increases the iteration count and wall time, although both alternatives remain convergent. The effect of using GAMG for the $\tilde{S}^{-1}$ action is more dependent on the accompanying $K^{-1}$ approximation. For the velocity-based formulation, Jacobi+GAMG converges more slowly than Jacobi+BoomerAMG, whereas BoomerAMG+GAMG fails to reach the prescribed tolerance within 2000 iterations. GAMG+GAMG, in contrast, converges in almost the same number of iterations but requires substantially more wall time than GAMG+BoomerAMG, indicating a higher cost per iteration. For the displacement-based formulation, all six configurations converge, but replacing BoomerAMG with GAMG for the $\tilde{S}^{-1}$ action consistently increases wall time; the deterioration is particularly pronounced for Jacobi+GAMG and GAMG+GAMG.

Overall, the results indicate that BoomerAMG provides the most effective approximation for both inverse actions in the present Block-Schur configuration, while also demonstrating that convergence rate and computational cost cannot be inferred from the choice of either sub-block solver in isolation.

\subsubsection{Performance of nested vs. non-nested preconditioners}

\begin{figure}[htbp]
    \centering
    \includegraphics[width=0.9\linewidth]{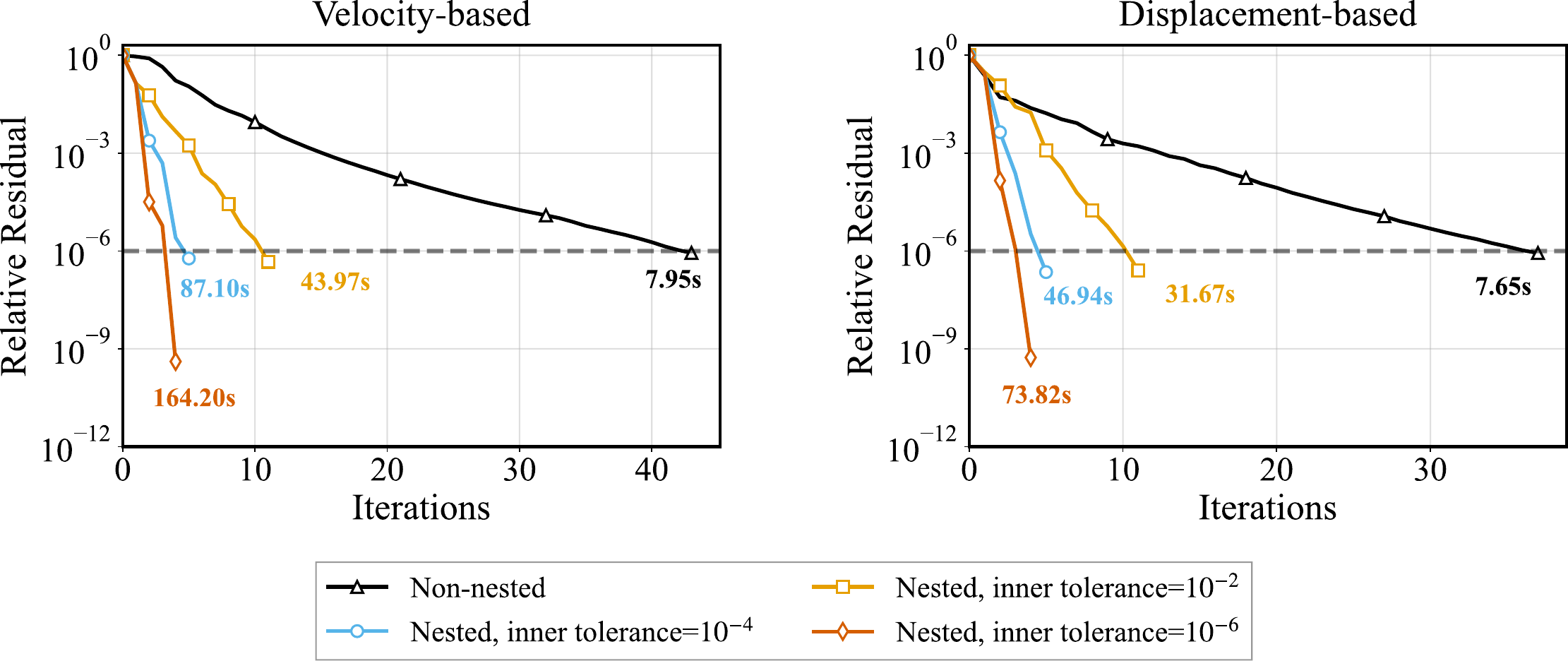}
    \caption{Performance comparison of nested and non-nested Schur-complement preconditioners, using the velocity-based (left) and displacement-based (right) structural formulations.}
    \label{fig:convergence-nested}
\end{figure}

We finally compare the non-nested Schur-complement preconditioner with the nested variant. For a fair comparison, both approaches employ a lower block-triangular factorization. In the nested block preconditioner, we set the maximum inner iterations $N_0=N_1=200$, and monitor the convergence performance against three different inner tolerances: $\varepsilon_{0}=\varepsilon_{1}=\varepsilon_{2}=10^{-2}, 10^{-4}, 10^{-6}$, following the parameter choices reported in literature \citep{LIU-nest,sun2025parallel}.

Figure~\ref{fig:convergence-nested} shows a clear trade-off between outer convergence and computational cost. For both structural formulations, tightening the inner tolerance from $10^{-2}$ to $10^{-6}$ progressively reduces the number of outer FGMRES iterations, with the tightest tolerance converging in only a few iterations. This improvement, however, comes at a substantial increase in wall time because each application of the nested preconditioner requires more expensive inner solves. Consequently, the non-nested Block-Schur configuration remains notably faster despite requiring considerably more outer iterations.

The reduced outer iteration count indicates that tighter inner solves provide a more effective preconditioning action, but the increased cost per iteration outweighs this benefit. Among the configurations considered, the non-nested Block-Schur preconditioner therefore provides the best overall computational efficiency for both structural formulations.

\section{Discussion} \label{sec:discussion}

The numerical results show that solver performance does not generalize consistently across the investigated cardiovascular FSI configurations. Changes in vascular geometry, physiological boundary conditions, flow characteristics, Krylov method, and structural formulation all affect solver behavior, with the strongest sensitivity observed for the structural formulation. Methods that perform robustly for the velocity-based formulation often fail for the displacement-based formulation. GMRES also generally exhibits more robust convergence than BiCGStab, consistent with its residual-minimizing property \citep{saad1986gmres}. Among the investigated preconditioners, Block-Schur provides the most consistent convergence across the full set of cases, whereas ML remains viable in most of the cases but generally at higher computational cost. The following discussion therefore focuses on the algebraic origin of the formulation-dependent behavior and on the features that contribute to the robustness of Schur-complement block preconditioning.

\subsection{Effect of structural formulation on block structure and solver robustness}

\begin{figure}[htbp]
    \centering
    \includegraphics[width=0.75\linewidth]{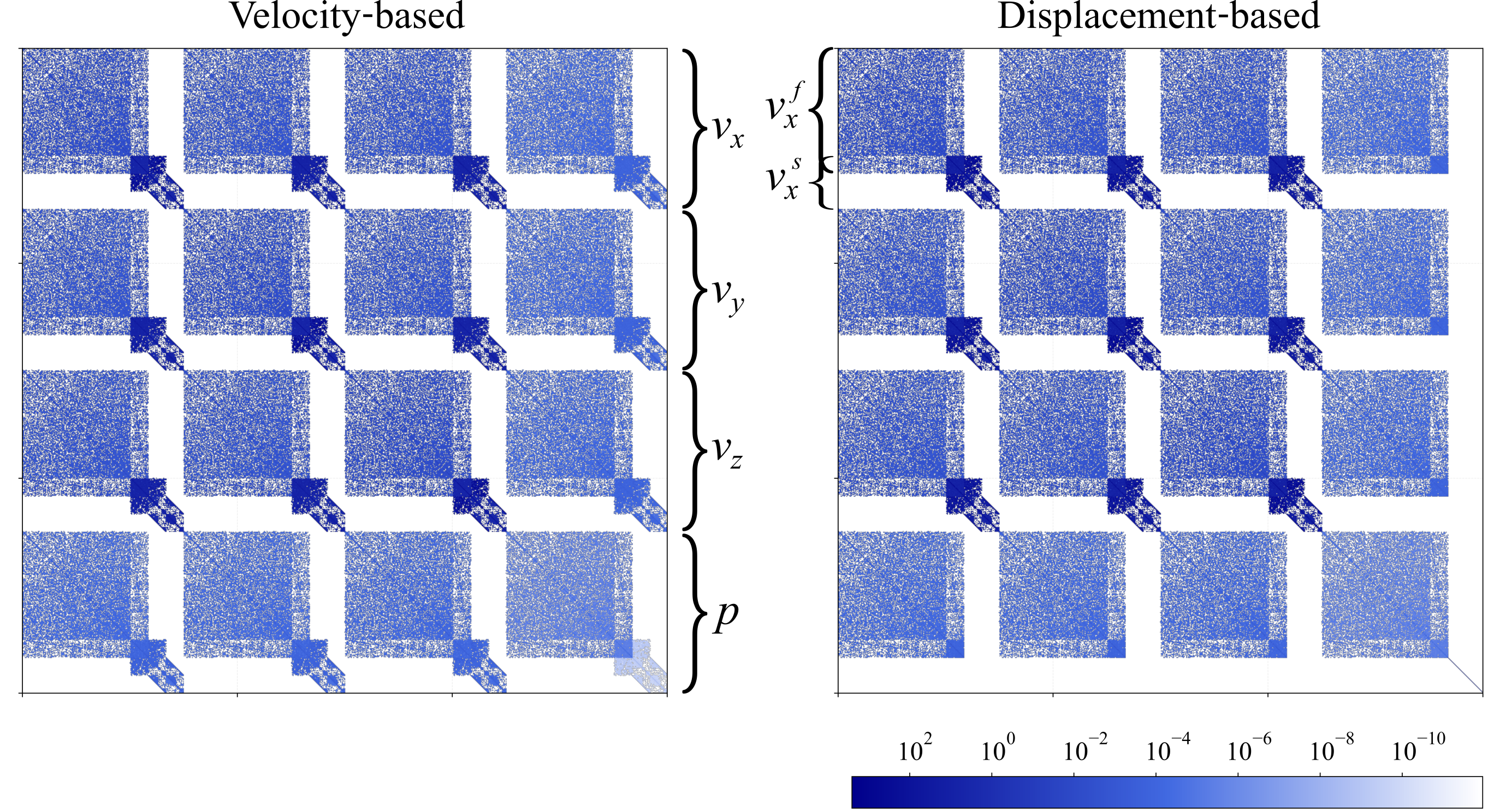}
    \caption{Sparsity patterns and entry magnitudes of the global FSI matrices for the velocity-based (left) and displacement-based (right) structural formulations.}
    \label{fig:mat-view}
\end{figure}

Across the cylindrical benchmark and the patient-specific cases, substantially fewer preconditioners converge with the displacement-based formulation. By contrast, among the methods that remain convergent under both formulations, the change in wall time is generally less pronounced. The formulation change therefore primarily affects the robustness of the investigated preconditioners.

To investigate the algebraic origin of this behavior, we compare the block structures and spectral properties of the linear systems generated with the two structural formulations. Since direct visualization of the production-scale matrices and computation of their condition numbers are computationally prohibitive, we use a uniformly coarsened aortic mesh containing approximately 1.3\% of the elements in the aortic mesh listed in Table~\ref{tab:patient_mesh}, while preserving the fluid-to-solid element ratio.

The change in structural formulation is reflected primarily in the pressure block $L$ and the coupling blocks $G$ and $D$, as shown in Figure~\ref{fig:mat-view}. In $L$, the velocity-based formulation contains solid-domain pressure contributions, including those arising from stabilization, whereas the displacement-based formulation has no physical pressure degrees of freedom in the solid domain. For implementation consistency, the corresponding variables are retained as dummy degrees of freedom in the displacement-based formulation, with unit diagonal entries and zero off-diagonal entries in the associated rows and columns. The structures of $G$ and $D$ also differ because the velocity-based formulation contains additional solid-domain velocity–pressure coupling. These differences modify the pressure-related and coupling structure of the monolithic system, while no apparent change is observed in the momentum block $K$ from the sparsity patterns.

\begin{figure}[htbp]
    \centering
    \includegraphics[width=0.7\linewidth]{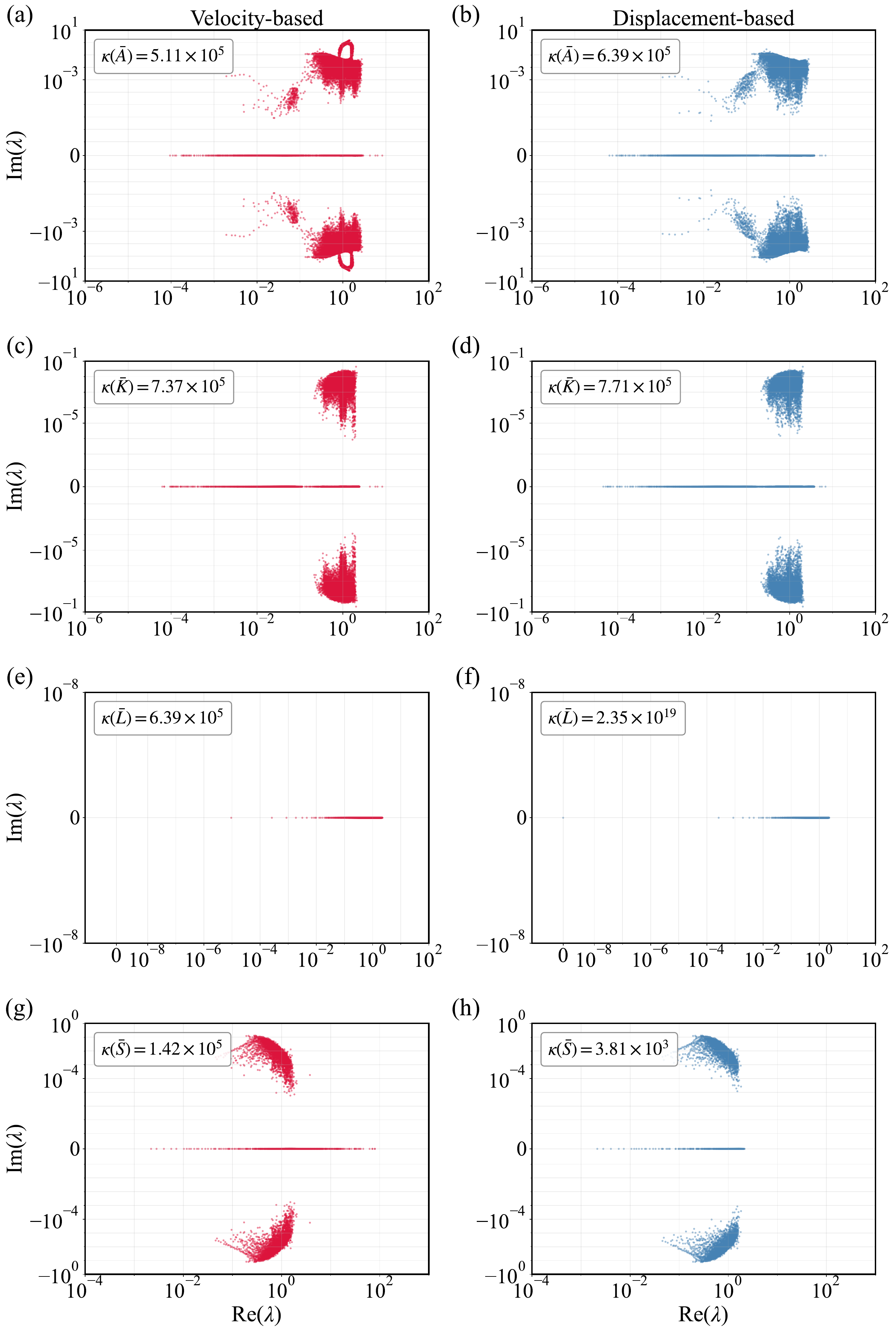}
    \caption{Eigenvalue distributions and condition numbers of the symmetrically Jacobi-scaled FSI system and its block components for the coarse aortic model at peak systole. The left and right columns correspond to the velocity- and displacement-based structural formulations, respectively. From top to bottom, the rows show the global system matrix, the velocity block, the pressure block, and the exact Schur complement. }
    \label{fig:scaled-eigs}
\end{figure}

The spectral analysis in Figure~\ref{fig:scaled-eigs} provides a block-level explanation for the strong formulation dependence observed in the solver results. To reduce the influence of differences in variable scaling, the matrices are symmetrically Jacobi scaled following Ref. \cite{Jongming1},
\begin{equation}
\bar{A}=Q^{-1/2}AQ^{-1/2},
\qquad
Q=\operatorname{diag}(A).
\end{equation}
The scaled global systems have comparable condition numbers, $5.11\times10^{5}$ and $6.39\times10^{5}$ for the velocity- and displacement-based formulations, respectively. The momentum blocks show similarly close values, $7.37\times10^{5}$ and $7.71\times10^{5}$, together with qualitatively similar eigenvalue distributions. Hence, the widespread solver failures observed for the displacement-based formulation cannot be attributed simply to poorer conditioning of the global matrix or the velocity block.

A much larger difference appears in the pressure block $L$. Its condition number is $6.39\times10^{5}$ for the velocity-based formulation but increases to $2.35\times10^{19}$ for the displacement-based formulation, with eigenvalues extending extremely close to zero.
The dummy degrees of freedom contribute unit singular values, which lie within the singular-value range of the physical pressure block and therefore do not alter its condition number. Direct computation after eliminating the dummy pressure variables confirms that the condition number is unchanged.
This distinction is particularly important for Block-Additive and Block-Multiplicative, both of which directly use the pressure block $L$ in the preconditioner. Its extremely poor conditioning in the displacement-based formulation makes the corresponding inverse action difficult to approximate effectively. This observation is consistent with the systematic failure of these preconditioners for the displacement-based cases.

The behavior changes fundamentally when the velocity-pressure coupling is retained through the Schur complement. For the displacement-based formulation, the condition number decreases from $2.35\times10^{19}$ for $L$ to only $3.81\times10^{3}$ for $S$. The coupling term $-DK^{-1}G$ therefore removes much of the severe ill-conditioning present in the isolated pressure block. In fact, the displacement-based Schur complement is better conditioned than its velocity-based counterpart, for which $\kappa(S)=1.42\times10^{5}$.
The same trend is observed for the approximate Schur complement $\tilde{S}$, whose condition numbers are $2.02\times10^{4}$ and $4.47\times10^{4}$ for the displacement-based and velocity-based formulations, respectively. Thus, despite the difference in their absolute condition numbers, both Schur complements exhibit better conditioning for the displacement-based formulation.
The difficulty of the displacement-based system is thus not intrinsic to the reduced pressure problem itself, but arises when the pressure block is treated independently of its coupling to the momentum equations. This provides an algebraic explanation for why Block-Schur remains effective when the simpler block splittings fail.

\subsection{Robustness and efficiency of Schur-complement block preconditioning}

As discussed in the previous section, Schur-complement block preconditioning exhibits the most consistent convergence across all investigated configurations because it better preserves the velocity-pressure coupling and improves the conditioning of the reduced pressure problem. Its practical efficiency, however, depends on how this coupling is incorporated and how the required block inverse actions are approximated.

The factorization study shows that retaining one off-diagonal coupling block captures most of the benefit of the full block factorization. This observation is consistent with established analyses of saddle-point systems, in which Schur-based triangular preconditioners can produce tightly clustered spectra without approximating the complete inverse of the coupled matrix \citep{Benzi_Golub_Liesen_2005}. In the present simulations, the full factorization provides only a small reduction in Krylov iterations, which is outweighed by the additional cost of applying the second coupling operator. The lower triangular factorization therefore offers the most favorable balance between coupling preservation and application cost.

The choice of algebraic multigrid method has a substantial influence on the Schur-complement solve. PETSc GAMG and BoomerAMG employ different coarsening, interpolation, and relaxation strategies, and their performance can depend strongly on the algebraic properties of the operator. The approximate Schur complement $\tilde{S}$ is generally nonsymmetric and differs between the two structural formulations, making it a nontrivial target for either AMG method. Under the configurations tested here, BoomerAMG provides a more effective approximation of the $\tilde{S}^{-1}$ action, whereas the corresponding GAMG configurations are generally less efficient and in some cases exhibit poor convergence. The present results do not identify which AMG component is responsible for this difference, and should therefore be interpreted as configuration specific rather than as evidence of the general superiority of BoomerAMG.

The nested comparison further demonstrates that increasing preconditioner accuracy does not necessarily improve computational efficiency. FGMRES permits the use of variable inner iterations \citep{saad1993flexible}, but tighter inner tolerances substantially increase the cost of each preconditioner application. In the present cases, the reduction in outer iterations is insufficient to offset this additional work, consistent with the general principle that the accuracy of inexact inner solves should be balanced against the convergence requirements of the outer Krylov method \citep{Simoncini2003}.

Overall, the results support a relatively simple Schur-complement design: retain one direction of the velocity-pressure coupling, use a robust approximation of the reduced pressure operator, and avoid oversolving the inner systems. Within the present framework, the lower triangular, non-nested Schur preconditioner with BoomerAMG approximations of both $K$ and $\tilde{S}$ provides the best balance between robustness and computational cost.

\subsection{Practical implications for cardiovascular FSI solver selection}

Within the present svFSI framework, GMRES combined with the non-nested lower triangular Block-Schur preconditioner and BoomerAMG for the sub-block inverse actions provides a practical default for cardiovascular FSI simulations. This configuration offers the most consistent convergence across the investigated geometries and structural formulations while maintaining competitive wall times. Other methods may be more efficient for individual problems, but their use should be supported by case-specific verification. Because svFSI is open source and used by a broad cardiovascular modeling community, such a robust default is particularly valuable for users without specialized expertise in iterative solvers. When both structural formulations are appropriate for the modeling objective, the velocity-based formulation offers a broader range of robust solver options in the present implementation.

\subsection{Limitations}

Although the cases considered here span a range of geometries, flow regimes, and physiological boundary conditions, they do not capture all cardiovascular FSI settings, such as systems coupled to closed-loop lumped-parameter network models \citep{Jongming1}. All patient-specific cases employ the same Neo-Hookean wall model and material parameters. More complex constitutive descriptions, including anisotropic or fiber-reinforced models, may modify the structural tangent operator and consequently affect solver behavior \citep{schussnig2024}.

These reported results and recommendations are specific to the svFSI implementation and numerical settings considered here and should be reassessed for substantially different models, discretizations, or solver configurations. Every algebraic multigrid method offers numerous tunable choices, including coarsening, interpolation, smoothing, and coarse-grid parameters; more extensive tuning could alter the relative performance of GAMG, BoomerAMG, and ML. Nevertheless, the consistency of the observed trends across the investigated geometries, structural formulations, and Krylov methods supports the broader finding that formulation-induced block structure strongly influences solver robustness.

\section{Conclusion}
\label{sec:conclusion}

This study systematically investigated Krylov solvers with algebraic and physics-based preconditioners for monolithic cardiovascular FSI across an idealized cylindrical benchmark and three patient-specific models. Solver behavior depends strongly on both the cardiovascular configuration and the structural formulation, and solver performance does not consistently transfer across the investigated cardiovascular FSI problems, including among the patient-specific cases. In particular, the displacement-based formulation substantially reduces the robustness of several preconditioners. Block-level spectral analysis shows that this behavior is associated with severe ill-conditioning of the isolated pressure block, whereas retaining the velocity-pressure coupling through the Schur complement greatly improves the conditioning of the reduced pressure problem.

Among the investigated approaches, Schur-complement block preconditioning provides the most consistent convergence across geometries and structural formulations. The focused comparisons further show that a relatively simple non-nested lower triangular Block-Schur configuration provides an effective balance between robustness and computational cost, with BoomerAMG giving the most reliable sub-block approximations under the settings considered here. These results provide a practical basis for linear-solver design in svFSI and highlight the importance of accounting for formulation-induced block structure when developing robust solvers for cardiovascular FSI.

\section*{CRediT authorship contribution statement}
\textbf{Zheyi Chu.} Methodology, Software, Formal analysis, Investigation, Writing - original draft, Writing - review \& editing. \textbf{Chi Zhu.} Conceptualization, Methodology, Software, Supervision, Funding acquisition, Writing - review \& editing.

\section*{Declaration of competing interest}
The authors declare that they have no known competing financial interests or personal relationships that could have appeared to influence the work reported in this paper.

\section*{Data availability}
Patient-specific models are available from Vascular Model Repository (\url{https://www.vascularmodel.com}).

\section*{Code availability}
The source code used in this study is openly available at \url{https://github.com/CZHU20/svFSI/}.

\section*{Acknowledgements}
This work was supported by the National Key Research and Development Program of China (Grant No. 2021YFA1000201), the National Natural Science Foundation of China (Grant Nos. 12293001, 12272009), and the Beijing Natural Science Foundation (Grant No. L259063). Software development also benefited from hardware support provided by the PKU Kunpeng-Ascend Scientific and Educational Innovation Excellence Center.

\section*{Declaration of generative AI and AI-assisted technologies in the manuscript preparation process}
The authors used Codex (OpenAI) to assist with language editing, manuscript organization, and checks of clarity and internal consistency. The authors evaluated the suggestions, revised the text, and verified the scientific content. The authors remain fully responsible for the final manuscript.

\bibliographystyle{unsrtnat}
\bibliography{references}

\end{document}